%% file: neutrinf.tex
\documentclass[a4paper,11pt]{article}
\pdfoutput=1 
\usepackage{jcappub}
\usepackage{amsmath}
\usepackage{graphicx}
\usepackage{latexsym}
\usepackage{xspace}
\usepackage{color}
\usepackage{hyperref} 
\usepackage{bm}
\usepackage{relsize}
\usepackage{tabularx}
\usepackage{multirow}
\usepackage{amssymb}
\usepackage[table]{xcolor}
\usepackage{slashbox}

\makeatletter
\gdef\@fpheader{}
\g@addto@macro\bfseries{\boldmath}
\makeatother

\input{newcommands}

\subheader{}

\title{Inflation Model Selection meets Dark Radiation}

\author[a]{Thomas Tram,}
\author[a]{Robert Vallance}
\author[a]{and Vincent Vennin}

\affiliation[a]{Institute of Cosmology \& Gravitation, University of Portsmouth, Dennis Sciama Building, Burnaby Road, Portsmouth, PO1 3FX, United Kingdom}

\emailAdd{thomas.tram@port.ac.uk}
\emailAdd{robert.vallance@student.manchester.ac.uk}
\emailAdd{vincent.vennin@port.ac.uk}

\date{today}

\begin{document}
\sloppy

\abstract{
We investigate how inflation model selection is affected by the presence of additional free-streaming relativistic degrees of freedom, \ie dark radiation. We perform a full Bayesian analysis of both inflation parameters and cosmological parameters taking reheating into account self-consistently. We compute the Bayesian evidence for a few representative inflation scenarios in both the standard $\Lambda$CDM model and an extension including dark radiation parametrised by its effective number of relativistic species $N_\ueff$. Using a minimal dataset (Planck low-$\ell$ polarisation, temperature power spectrum and lensing reconstruction), we find that the observational status of most inflationary models is unchanged. The exceptions are potentials such as power-law inflation that predict large values for the scalar spectral index that can only be realised when $N_\ueff$ is allowed to vary. Adding baryon acoustic oscillations data and the B-mode data from BICEP2/Keck makes power-law inflation disfavoured, while adding local measurements of the Hubble constant $H_0$ makes power-law inflation slightly favoured compared to the best single-field plateau potentials. This illustrates how the dark radiation solution to the $H_0$ tension would have deep consequences for inflation model selection.}


\arxivnumber{1606.09199}

\maketitle

\section{Introduction}
\label{sec:intro}
Inflation~\cite{Starobinsky:1980te, Sato:1980yn, Guth:1980zm, Linde:1981mu, Albrecht:1982wi, Linde:1983gd} is the leading paradigm to describe the physical conditions that prevailed in the very early Universe. During this accelerated expansion epoch, vacuum quantum fluctuations of the gravitational and matter fields were amplified to large-scale cosmological perturbations~\cite{Starobinsky:1979ty, Mukhanov:1981xt, Hawking:1982cz,  Starobinsky:1982ee, Guth:1982ec, Bardeen:1983qw}, that later seeded the Cosmic Microwave Background (CMB) anisotropies and the large scale structure of our Universe.

Recent high-quality measurements~\cite{Ade:2013sjv, Adam:2015rua, Array:2015xqh, Ade:2015lrj} of the CMB temperature and polarisation inhomogeneities have significantly improved our knowledge of inflation. At present, the full set of observations can be accounted for in a minimal setup, where inflation is driven by a single scalar field $\phi$ with canonical kinetic term, minimally coupled to gravity, and evolving in a flat potential $V(\phi)$ in the slow-roll regime. Since particle physics beyond the electroweak scale remains elusive, and given that inflation can proceed at energy scales as large as $10^{16}\GeV$, even within this class of models, hundreds of inflationary scenarios have been proposed. A systematic Bayesian analysis~\cite{Martin:2013tda, Martin:2013nzq, Vennin:2015eaa} reveals that one third of them can now be considered as ruled out, while the vast majority of the preferred scenarios are of the plateau type, \ie they are such that the potential $V(\phi)$ is a monotonic function that asymptotes a constant value when $\phi$ goes to infinity. 

Inflation also needs to be connected to the subsequent hot Big Bang phase through an era of reheating, during which the energy contained in the inflationary fields eventually decays into the standard model degrees of freedom. The amount of expansion during this epoch determines the amount of expansion between the Hubble crossing time of the physical scales probed in the CMB and the end of inflation~\cite{Martin:2006rs, Martin:2010kz, Easther:2011yq, Dai:2014jja, Rehagen:2015zma}. As a consequence, the kinematics of reheating sets the time frame during which the fluctuations probed in cosmological experiments emerge, hence defining the location of the observable window along the inflationary potential. This effect can be used to extract constraints on a certain combination of the averaged equation-of-state parameter during reheating and the reheating temperature~\cite{Martin:2014nya, Martin:2016oyk}.

The systematic analyses mentioned above all assume the standard cosmological model. However, one may wonder whether their conclusions are robust against extensions of the standard cosmological model. One such extension is to include an additional component of free-streaming relativistic particles, often called dark radiation. There are many candidates for dark radiation, including eV-scale sterile neutrinos~\cite{Archidiacono:2016kkh}, thermal axions~\cite{DiValentino:2013qma} and Goldstone bosons~\cite{Weinberg:2013kea}. It is also possible that dark matter is just one of the particles in an extended dark sector that also contains relativistic particles. Of all the candidates, the eV-scale sterile neutrino is arguably the best motivated, since its existence would explain several anomalies observed in laboratory experiments, see e.g. ref.~\cite{Gariazzo:2015rra} for a review.

The effect of dark radiation on the CMB anisotropy power spectrum is to increase the damping at large multipoles. This damping is induced by enhanced photon diffusion at the time of last scattering when the number of relativistic species is increased while the peak scale is kept fixed~\cite{Hou:2011ec}. To some extent it can be compensated by an increase in the tilt of the primordial scalar power spectrum, thus we expect dark radiation to affect inflation model selection. In this paper we investigate this effect assuming dark radiation is free-streaming, but a strongly coupled component would yield the same suppression while being less constrained by CMB data~\cite{Baumann:2015rya}.
 
The paper is organised as follows. In \Sec{sec:method} we present the inflationary models we consider and explain how they can be compared using Bayesian model selection techniques. In \Sec{sec:results} we show the ranking of the models and discuss its dependence on various combinations of data sets. The constraints on the reheating parameter, the effective number of neutrino species and the present expansion rate are also analysed. We finally present our conclusions in \Sec{sec:Conclusions}.
\section{Method}
\label{sec:method}
\begin{figure}[t]
\begin{center}
\includegraphics[width=\textwidth]{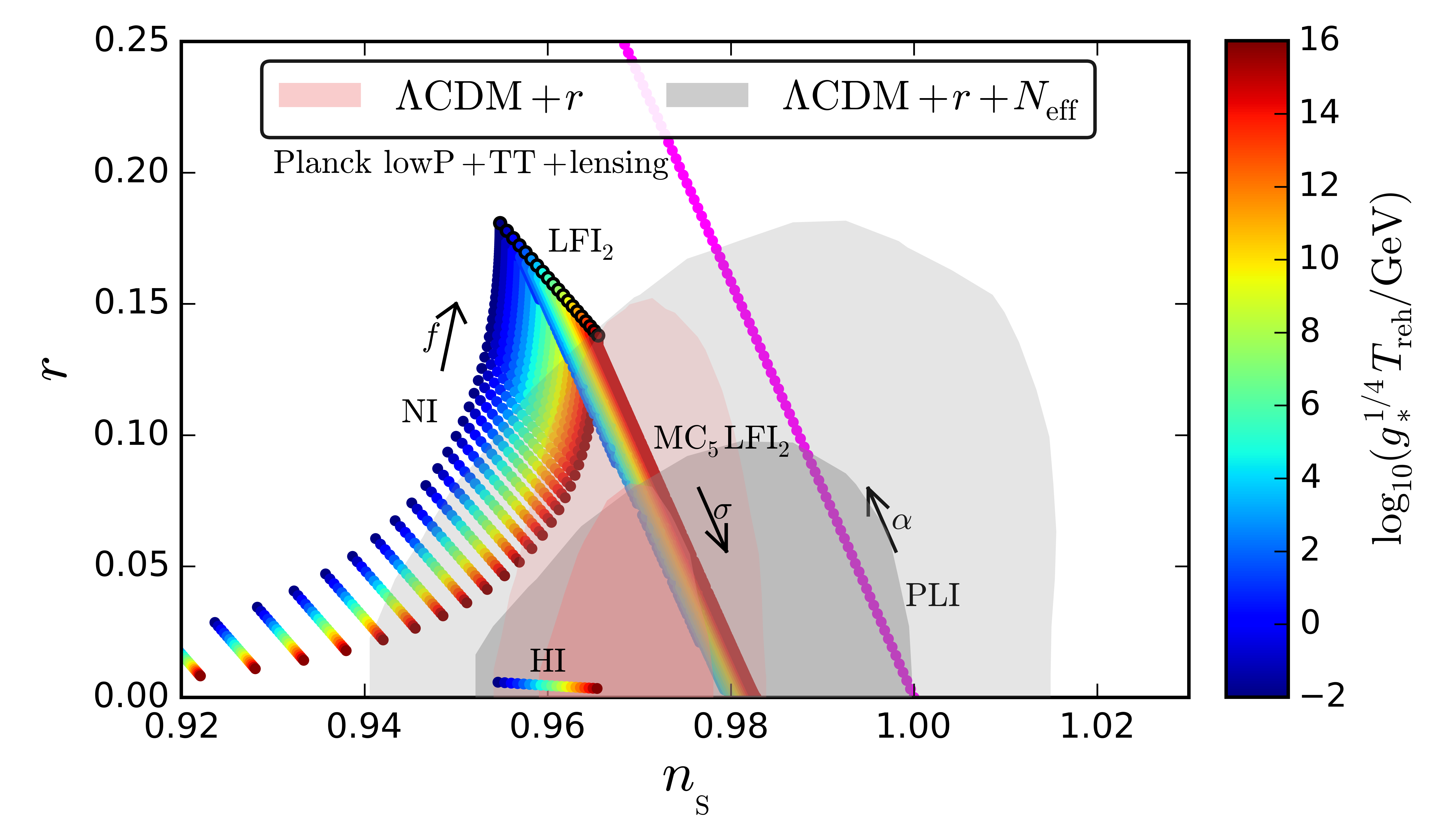}
\caption{Scalar spectral index $\nS$ and tensor-to-scalar ratio $r$ predicted by some of the models considered in this work. For natural inflation ($\mathrm{NI}$) and power-law inflation ($\mathrm{PLI}$), the arrow indicates in which direction $f$ [cf. \Eq{eq:pot:ni}] and $\alpha$ [cf. \Eq{eq:pot:pli}] respectively vary. For $\mathrm{MC}_5\mathrm{LFI}_2$ (see \Sec{sec:curvaton}), the arrow denotes how $\nS$ and $r$ change when the contribution from the curvaton field $\sigma$ increases. The two remaining models are quadratic large field inflation ($\mathrm{LFI}_2$) and Higgs inflation (the Starobinsky model, $\mathrm{HI}$). In this figure, the inflaton is assumed to oscillate at the quadratic minimum of its potential once inflation ends (this assumption is dropped in the rest of the paper and is used here for display convenience only) and the colour denotes the reheating temperature $T_\ureh$, see \Sec{sec:reheating} (for power-law inflation, the predictions are independent of $T_\ureh$). The pink shaded surfaces are the one- and two-sigma contours of the Planck 2015 lowP+TT+lensing data when the standard $\Lambda\mathrm{CDM}$ cosmological model is assumed, while for the grey shaded surfaces, the effective number of relativistic degrees of freedom $N_\ueff$ is allowed to vary.}
\label{fig:nsr}
\end{center}
\end{figure}
\subsection{Dark radiation and the primordial power spectrum}
\label{sec:darkRadiation_tilt}
As mentioned in \Sec{sec:intro}, dark radiation leads to extra suppression in the CMB anisotropy power spectra at small scales. This can be compensated by increasing the value of the spectral index $\nS\equiv \dd\ln\calP_\zeta/\dd \ln k$ of the primordial scalar power spectrum $\calP_\zeta$. This effect slightly depends on the amount of tensor perturbations, characterised by the tensor-to-scalar ratio $r_{0.05}=\calP_h/\calP_\zeta$ evaluated at the pivot scale $k_{{}_\mathrm{P}}=0.05\,\Mpc^{-1}$. In \Fig{fig:nsr} we have shown the one- and two-sigma contours in the $(\nS,r)$ plane derived from the Planck 2015 data~\cite{Aghanim:2015xee}, assuming the standard $\Lambda\mathrm{CDM}$ cosmological model (pink) and allowing for the effective number of neutrino species $N_\ueff$ to vary (grey). Here, $N_\ueff$ is defined so that the total relativistic energy density contained in neutrinos and any other dark radiation is given in terms of the photon density $\rho_\gamma$ at $T\ll 1$ MeV by
\bea
\rho = N_\ueff \frac{7}{8} \left(\frac{4}{11}\right)^{4/3} \rho_\gamma\, .
\eea
In the standard cosmological model where the only contributions to dark radiation are the three active neutrinos we have $N_\ueff \simeq 3.046$. This number differs slightly from 3 due to the details of the freeze-out process of the active neutrinos~\cite{Mangano:2001iu,deSalas:2016ztq}. When $N_\ueff$ is free to depart from this standard value, one can check in \Fig{fig:nsr} that larger values of $\nS$ are allowed as expected.
\subsection{Primordial power spectrum from inflation}
\label{sec:primordialPowerSpectrum_Inflation}
In what follows, unless otherwise specified, we examine the case where inflation is realised by a single scalar field $\phi$ slowly rolling down its potential $V(\phi)$. Instead of considering the hundreds of such models that have been proposed in the literature and recently listed in \Ref{Martin:2013tda}, we study a few prototypical examples. As a rule of thumb, models are disfavoured when they predict too large a value for the tensor-to-scalar ratio $r$, or if the scalar spectral index $\nS$ is either too small or too large. A typical example which predicts a too large value for $r$ is when the inflaton field $\phi$ is simply a massive field,
\bea
\label{eq:pot:lfi2}
V(\phi) = M^4\left(\frac{\phi}{\Mp}\right)^2\, .
\eea
This model is called $\mathrm{LFI}_2$ (for large-field inflation) in the terminology of \Ref{Martin:2013tda}, where $\Mp$ is the reduced Planck mass and $M$ is an overall mass scale. In natural inflation, $\mathrm{NI}$, the potential is given by
\bea
\label{eq:pot:ni}
V(\phi) = M^4\left[1+\cos\left(\frac{\phi}{f}\right)\right]
\eea
and yields values for $\nS$ that are too small when $f$ is not much larger than the Planck mass. Conversely, in power-law inflation, $\mathrm{PLI}$, the potential is of the form
\bea
\label{eq:pot:pli}
V(\phi) = M^4 \exp\left(-\alpha\frac{\phi}{\Mp}\right)
\eea
and the value predicted for $\nS$ is too large. Favoured potentials on the other hand are mostly of the plateau type. A typical example is Higgs inflation
\bea
V(\phi) = M^4 \left[1-\exp\left(-\sqrt{\frac{2}{3}}\frac{\phi}{\Mp}\right)\right]^2\, ,
\eea
also known as the Starobinsky model. The predictions of these potentials are shown in \Fig{fig:nsr}, under the assumption that the averaged equation-of-state parameter during reheating vanishes. This assumption is made for illustration purposes only and is dropped in what follows. To further study how the detailed ranking of the best inflationary models depends on the assumptions made about $N_\ueff$, two other plateau potentials are also included in the analysis, K\"ahler moduli I inflation (KMII) for which $V(\phi) = M^4(1-\alpha\phi/\Mp\ee^{-\phi/\Mp})$ and exponential SUSY inflation ($\mathrm{ESI}_\mathrm{o}$) for which $V(\phi) = M^4(1-\ee^{-q\phi/\Mp})$.

For all these potentials, the values of the slow-roll parameters at Hubble exit time of the pivot scale $k_{{}_\mathrm{P}}$ are computed using the \textsc{aspic} library~\cite{aspic}. The primordial scalar and tensor power spectra are then evaluated at second order in slow roll~\cite{Leach:2002ar, Choe:2004zg, Martin:2013uma}, and evolved using the Boltzmann code \textsc{class}~\cite{2011JCAP...07..034B}, which returns the CMB anisotropy power spectra. Their likelihood is then computed by \textsc{Clik} based on the Planck 2015 data~\cite{Aghanim:2015xee}. We include the low-$\ell$ polarisation, the temperature power spectrum and the lensing reconstruction. We perform the sampling using \textsc{MontePython}~\cite{2013JCAP...02..001A} in the nested sampling mode which relies on \textsc{MultiNest}~\cite{Feroz:2007kg,Feroz:2008xx,Feroz:2013hea} and \textsc{PyMultiNest}~\cite{Buchner:2014nha}.
\subsection{Bayesian model selection}
\label{sec:BayesianSelection}
This numerical pipeline returns the Bayesian evidence~\cite{Cox:1946,Jeffreys:1961,deFinetti:1974,Trotta:2008qt} $\mathcal{E}$ of the inflationary models $\mathcal{M}_i$ listed above,
\bea
\label{eq:evidence:def}
\mathcal{E}\left(\mathcal{D}\vert\mathcal{M}_i \right) 
= \int\dd\theta_{ij}\mathcal{L}
\left(\mathcal{D}\vert\theta_{ij},\mathcal{M}_i\right)
\pi\left(\theta_{ij}\vert \mathcal{M}_i\right)\, .
\eea
In this expression, $\theta_{ij}$ are the parameters defining the model $\mathcal{M}_i$ and the likelihood function $\mathcal{L}\left(\mathcal{D}\vert\theta_{ij},\mathcal{M}_i\right)$ represents the probability of observing the data $\mathcal{D}$ assuming the model $\mathcal{M}_i$ is true and $\theta_{ij}$ are the actual values of its parameters. The prior distribution $\pi\left(\theta_{ij}\vert \mathcal{M}_i\right)$ encodes the information one has a priori on the values of the parameters $\theta_{ij}$ describing the model. For the parameters of the potentials as well as the reheating parameter (see \Sec{sec:reheating}), we use the same priors as the ones proposed in \Ref{Martin:2013nzq}, which are based on the model-building considerations of \Ref{Martin:2013tda}. We consider two scenarios for the dark radiation sector, both of them having a massive neutrino of minimum mass  $0.06\, \mathrm{eV}$. In the first one, two massless and one massive neutrino are present, and $N_\ueff$ takes its standard value $N_\ueff=3.046$. In the second case, extra massless and free-streaming dark radiation components are allowed and $\Delta N_\ueff \equiv N_\ueff - 3.046$ varies in the range $[-2,3]$ (corresponding to taking the number of ultrarelativistic species $N_\mathrm{ur}\in[0,5]$) with a flat prior. Otherwise, all priors for the cosmological parameters have been kept identical to those used by the Planck collaboration, see table 4 in \Ref{Ade:2013zuv}. 

Under the principle of indifference, two models $\mathcal{M}_i$ and $\mathcal{M}_j$ can be compared by computing the ratio $\mathcal{E}_i/\mathcal{E}_j$ of their Bayesian evidence. This ratio is called the Bayes factor and one can interpret this number using Jeffreys' empirical scale in the following way. When $\ln(\mathcal{E}_i/\mathcal{E}_j)>5$, $\mathcal{M}_j$ is said to be ``strongly disfavoured'' with respect to $\mathcal{M}_i$, ``moderately disfavoured'' if $2.5<\ln(\mathcal{E}_i/\mathcal{E}_j)<5$, ``weakly disfavoured'' if $1<\ln(\mathcal{E}_i/\mathcal{E}_j)<2.5$, and the result is said to be ``inconclusive'' if $\vert \ln(\mathcal{E}_i/\mathcal{E}_j)\vert<1$.
\section{Results}
\label{sec:results}
\begin{figure}[t]
\begin{center}
\includegraphics[width=\textwidth]{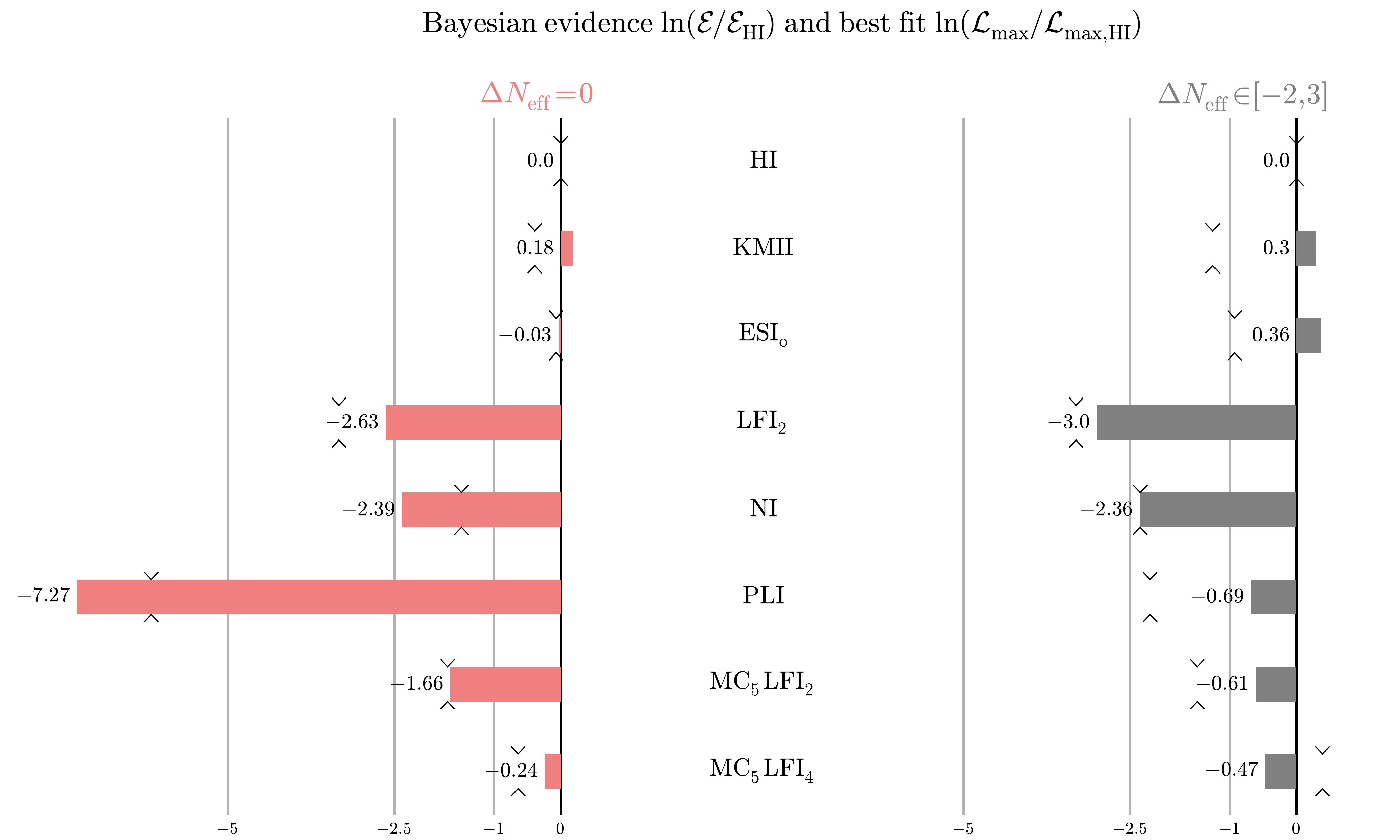}
\caption{Bayesian evidence of the inflationary models considered in this work when $N_\ueff = 3.046$ is fixed to its standard value (left column, pink) and is allowed to vary in the interval $\Delta N_\ueff\in[-2,3]$ (right column, grey). In both cases, the Bayesian evidence is normalised to Higgs inflation (HI, the Starobinsky model), taken as a reference model. The analysis is performed with the Planck 2015 lowP + TT + lensing data. The typical numerical sampling error is around $\ln\mathcal{E}\sim 0.3$. The vertical grey lines denote Jeffreys' scale, and the ${}^\vee_\wedge$ symbols display the best-fit values normalised to HI.}
\label{fig:evid}
\end{center}
\end{figure}
\subsection{Inflationary model selection}
\label{sec:modelSelection}
In \Fig{fig:evid}, the Bayesian evidences of the inflationary models listed in \Sec{sec:method} are displayed, when $N_\ueff = 3.046$ is fixed to its standard value (left column) and when $\Delta N_\ueff = N_\ueff - 3.046$ is allowed to vary in the interval $[-2,3]$ (right column). The Bayes factors between the three plateau potentials, $\mathrm{HI}$, $\mathrm{KMII}$ and $\mathrm{ESI}_\mathrm{o}$, are rather insensitive to changing the assumption on $N_\ueff$, at the ``inconclusive'' level according to Jeffreys' scale. This means that the ranking amongst the best single-field slow-roll models is rather robust under the introduction of extra dark radiation components. 

For $\mathrm{LFI}_2$, a prototypical example yielding a value for $r$ that is too large, one can see that the model remains disfavoured when $N_\ueff$ is allowed to vary, and its Bayesian evidence compared to the best plateau potentials even decreases slightly. This may seem counter-intuitive since $r$ can a priori be made larger when $N_\ueff$ departs from its standard prediction. However, as we can see from \Fig{fig:nsr}, larger values of $r$ imply larger values of $\nS$ as well, that $\mathrm{LFI}_2$ cannot accommodate. This explains why large-field models cannot fit the data, even in this extended cosmology. One may also wonder why $\mathrm{LFI}_2$ is not found to be as disfavoured as usually claimed in the literature~\cite{Ade:2015lrj,Martin:2013nzq}. The reason is that here we consider a conservative data set made of low-$\ell$ polarisation, temperature power spectrum (TT) and lensing reconstruction (the inclusion of other data sets are discussed in \Secs{sec:datasets} and~\ref{sec:H_0}). Including other measurements (PlanckTT+lowP+BAO in \Ref{Ade:2015lrj} and PlanckTT,TE,EE+lowTEB~\cite{Aghanim:2015xee}+BICEP2-Keck/Planck likelihood~\cite{Ade:2015tva} in \Ref{Martin:2013nzq}) yields tighter upper bounds on $r$, and hence firmer exclusion of $\mathrm{LFI}_2$. This makes our conclusion all the more robust, namely that $\mathrm{LFI}_2$ is disfavoured with or without extra relativistic species.

The Bayesian evidence of natural inflation ($\mathrm{NI}$) is also almost unchanged, and the model remains weakly disfavoured compared to $\mathrm{HI}$, and even moderately disfavoured compared to the best model. This is because, when $f\lesssim \Mp$, the model predicts values for $\nS$ that are too small whether or not $N_\ueff$ is fixed. Indeed, we obtain the $2$-$\sigma$ constraint $\log(f/\Mp)>0.72$ when $\Delta N_\ueff=0$ and $\log(f/\Mp)>0.62$ when $N_\ueff$ is allowed to vary. On the other hand, when $f\gg \Mp$, the model asymptotes $\mathrm{LFI}_2$, which as we already discussed remains disfavoured. In between, there is an intermediate range of values for $f$ where one can see in \Fig{fig:nsr} that the model may accommodate the data, but it is so fine-tuned that it does not lead to a substantial increase in the Bayesian evidence. 

The only model for which a significant change in the Bayesian evidence occurs is power-law inflation ($\mathrm{PLI}$), which is strongly disfavoured in the standard case but falls in the inconclusive zone of $\mathrm{HI}$ when $N_\ueff$ is allowed to vary (and is only weakly disfavoured compared to the best model). This is because the values for $\nS$ predicted by $\mathrm{PLI}$, otherwise too large to fit the data in the standard case, are allowed when $N_\ueff$ varies. The status of this model is therefore strongly dependent on the assumptions made about the dark radiation sector of the underlying cosmology. Let us also notice that when $\Delta N_\ueff=0$, one obtains the $1$-$\sigma$ constraint $\log\alpha=-0.895_{-0.078}^{+0.073}$, but the model is so disfavoured in this case that this should not be interpreted as a measurement. If $N_\ueff$ is allowed to vary, the $2$-$\sigma$ upper bound $\log\alpha<-1.062$ is obtained.

In summary, the standard ranking of inflationary models is mostly robust under the introduction of extra dark radiation components: plateau potentials remain favoured while large-field potentials and models yielding values for $\nS$ that are too small such as natural inflation remain disfavoured. The only exception concerns models that predict values for $\nS$ that are too large in the standard cosmology, such as power-law inflation, which can become favoured once $N_\ueff$ is allowed to depart from its standard value. 
\subsection{An example of non single-field slow-roll models: curvaton scenarios}
\label{sec:curvaton}
The fact that the data can be accounted for in the simplest framework where inflation is driven by a single scalar field in the slow-roll regime does not necessarily mean that more complicated models are ruled out. This is why in this section, we illustrate the consequences of allowing $N_\ueff$ to vary on non-standard scenarios of inflation with the example of curvaton scenarios~\cite{Linde:1996gt, Enqvist:2001zp, Lyth:2001nq}.

In this setup, a light scalar field that behaves as a pure spectator field during inflation can dominate the energy budget of the Universe afterwards, for instance if it decays after the inflaton. It then contributes to scalar curvature perturbations and changes its power spectrum and non-Gaussianity. In the standard cosmological scenario, it has recently been shown~\cite{Vennin:2015egh} that the favoured curvaton models are of two kinds: either the inflaton potential is of the plateau type, or it has a quartic profile $V(\phi)\propto\phi^4$ and the curvatonic reheating scenario must be of the $5^\mathrm{th}$ or $8^\mathrm{th}$ type (according to the classification of~\Ref{Vennin:2015vfa}). 

In particular, curvaton models with a quadratic inflaton potential are disfavoured since, even if they can give rise to a value for $r$ that is small enough, it is at the expense of $\nS$ being too large. This case is displayed in \Fig{fig:nsr} and labeled as $\mathrm{MC}_5\mathrm{LFI}_2$ where $\mathrm{MC}_5$ means ``massive curvaton in the $5^\mathrm{th}$ scenario'' and $\mathrm{LFI}_2$ refers to the quadratic inflaton potential. Since larger values of $\nS$ are allowed when $N_\ueff$ is free to depart from its standard value, the status of curvaton models may a priori change. We computed the Bayesian evidence for two large-field curvaton models $\mathrm{MC}_5\mathrm{LFI}_2$ and $\mathrm{MC}_5\mathrm{LFI}_4$, where the inflaton potential is respectively quadratic and quartic and the curvaton scenario is of the $5^\mathrm{th}$ type. 

The resulting evidences are shown in \Fig{fig:evid}. In the standard cosmological model, one can see that $\mathrm{MC}_5\mathrm{LFI}_2$ is indeed disfavoured compared to $\mathrm{MC}_5\mathrm{LFI}_4$, that has a similar Bayesian evidence as the best single-field plateau potentials. When $N_\ueff$ is allowed to vary however, the two models have a comparable Bayesian evidence, and are both favoured. Similarly to $\mathrm{PLI}$, the observational status of $\mathrm{MC}_5\mathrm{LFI}_2$ is therefore dependent on the assumptions made about the dark radiation sector. 
\subsection{Combining different data sets}
\label{sec:datasets}
\begin{figure}[t]
\begin{center}
\includegraphics[width=0.49\textwidth]{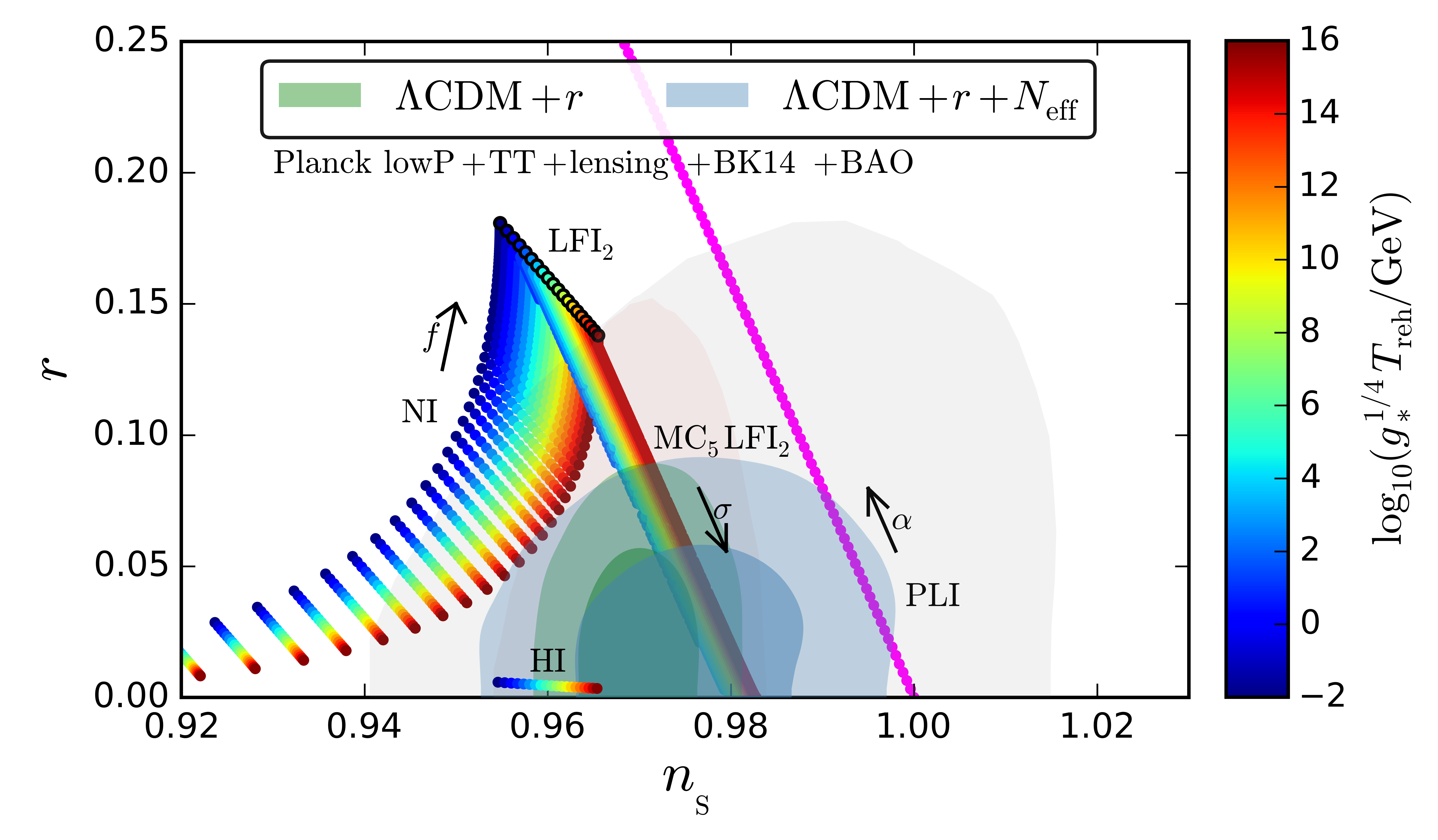}
\includegraphics[width=0.49\textwidth]{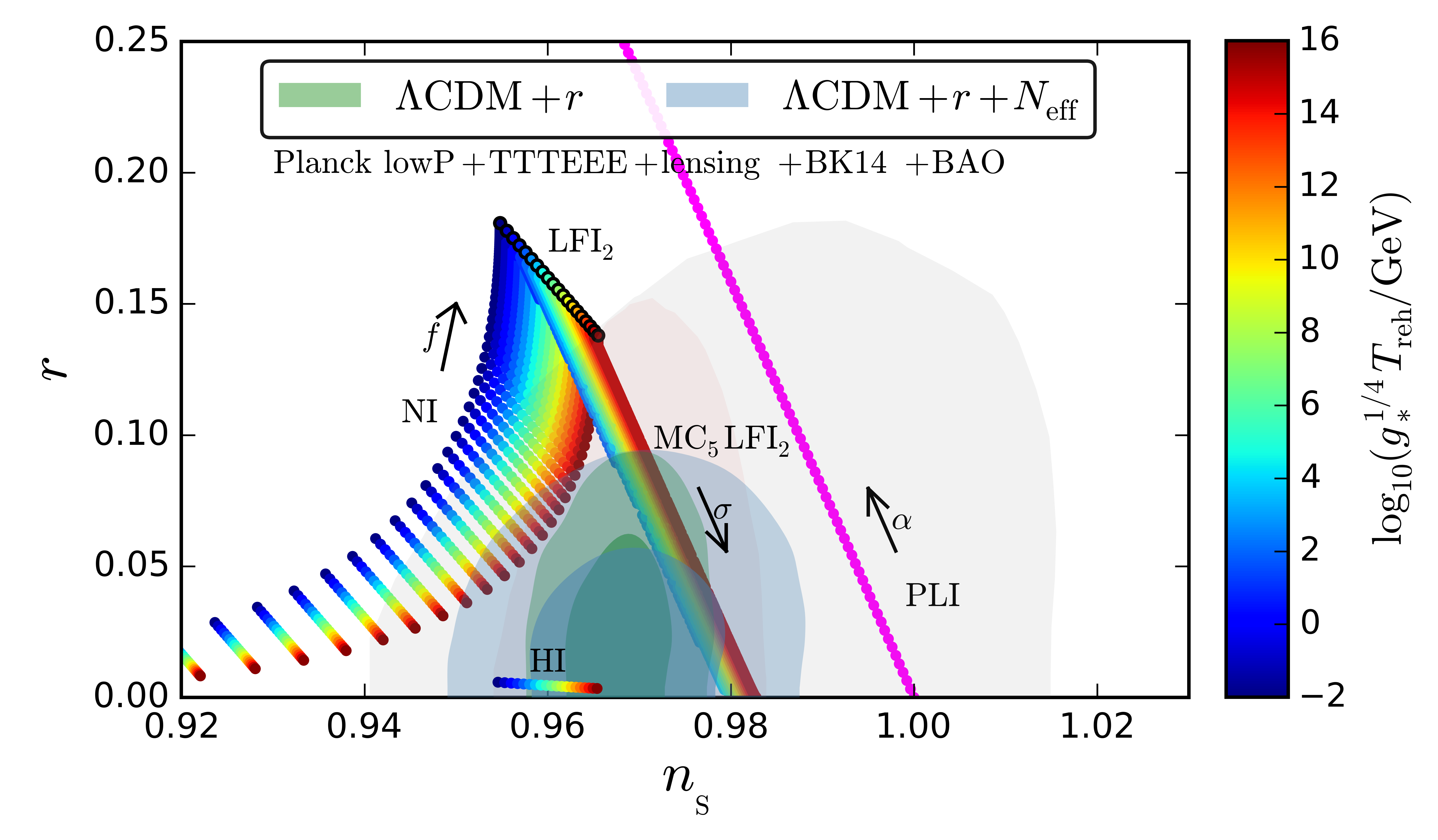}
\caption{Same as in \Fig{fig:nsr} with observational constraints derived when adding the Bicep2/Keck (BK14) and BAO data in the left panel, and the Bicep2/Keck + BAO + Planck high-$\ell$ polarisation data (TTTEEE) in the right panel. The green shaded surfaces are the one and two sigma contours when the standard $\Lambda\mathrm{CDM}$ cosmological model is assumed, while for the blue shaded surfaces, the effective number of relativistic degrees of freedom $N_\ueff$ is allowed to vary. In both panels, the constraints from the base Planck 2015 lowP+TT+lensing data displayed in \Fig{fig:nsr} have been recalled (pink and grey shaded areas).}
\label{fig:nsr:datasets}
\end{center}
\end{figure}
So far the constraints on inflationary models have been discussed using a minimal dataset consisting of Planck low-$\ell$ polarisation, temperature power spectrum and lensing reconstruction. In this section, we study how these results change when other data sets are included in the analysis. 

In the left panel of \Fig{fig:nsr:datasets}, the BICEP2/Keck (BK14) + Baryon Acoustic Oscillations (BAO) data~\cite{Anderson:2013zyy, Array:2015xqh, Ade:2015tva} has been added when deriving the posterior distribution on $\nS$ and $r$. One can see that, compared to the minimal dataset used in \Fig{fig:nsr}, the upper bound on $r$ is decreased, and values of $\nS$ larger than one are now disfavoured at the 2-$\sigma$ level when $N_\ueff$ is allowed to vary. This leaves the Bayesian evidence of the inflationary models discussed in \Fig{fig:evid} mostly unchanged except for $\mathrm{LFI}_2$ and $\mathrm{PLI}$. For $\mathrm{LFI}_2$, as explained in \Sec{sec:modelSelection}, the model becomes strongly disfavoured regardless of whether $N_\ueff$ is allowed to vary or not, and the conclusions drawn previously remain unchanged. For $\mathrm{PLI}$ however, since the upper bounds on both $\nS$ and $r$ are decreased, the model becomes strongly disfavoured even when $N_\ueff$ is allowed to vary since one finds $\ln\left(\mathcal{E}^\mathrm{PLI}/\mathcal{E}^\mathrm{HI}\right)= -21.71$ when $N_\ueff$ is fixed to its standard value, and $\ln\left(\mathcal{E}^\mathrm{PLI}/\mathcal{E}^\mathrm{HI}\right)= -6.48$ when $N_\ueff$ is allowed to vary. Therefore, even though additional dark radiation still improves $\mathrm{PLI}$ significantly, it is not at a level where the model can be made favoured when the Bicep2/Keck + BAO data is included.

In the right panel of \Fig{fig:nsr:datasets}, the Planck high-$\ell$ polarisation data (TTTEEE) has also been added together with Bicep2/Keck and BAO. The constraints on $r$ are unchanged but one can see that the upper constraints on $\nS$ are made tighter, even when $N_\ueff$ is allowed to vary. This is because the high-$\ell$ polarisation data prevents large departures of $N_\ueff$ from its standard value. However, let us mention that the role played by systematics in the high-$\ell$ polarisation data still needs to be confirmed~\cite{Ade:2015xua}. Moreover, if dark radiation is made of a fluid, either tightly coupled or simply relativistic, the bound from polarisation becomes looser, $N_\ueff$ is allowed to be larger and so is $\nS$ (see \eg table 1 of~\Ref{Baumann:2015rya}). The constraints obtained in this case should therefore be interpreted carefully.

Let us finally notice that the Planck collaboration has recently performed an improved low-$\ell$ polarisation analysis~\cite{Aghanim:2016yuo} which shifts the optical depth $\tau$ to slightly smaller values. However, the new likelihood \textsc{SimLow} is not publicly available yet, which is why it is not included in our analysis.\footnote{A partial solution~\cite{DiValentino:2016ucb} is to use the posterior on $\tau$ derived by Planck in $\Lambda$CDM as a prior in both cosmologies ($\Lambda$CDM+$r$ and $\Lambda$CDM+$r$+$N_\ueff$). However, because $\tau$ is correlated with $\nS$, this is not fully consistent and this biases the result towards smaller values of $N_\ueff$. This is why we do not follow this approach.}

\subsection{Constraining the reheating parameter}
\label{sec:reheating}
\begin{figure}[t]
\begin{center}
\includegraphics[width=0.7\textwidth]{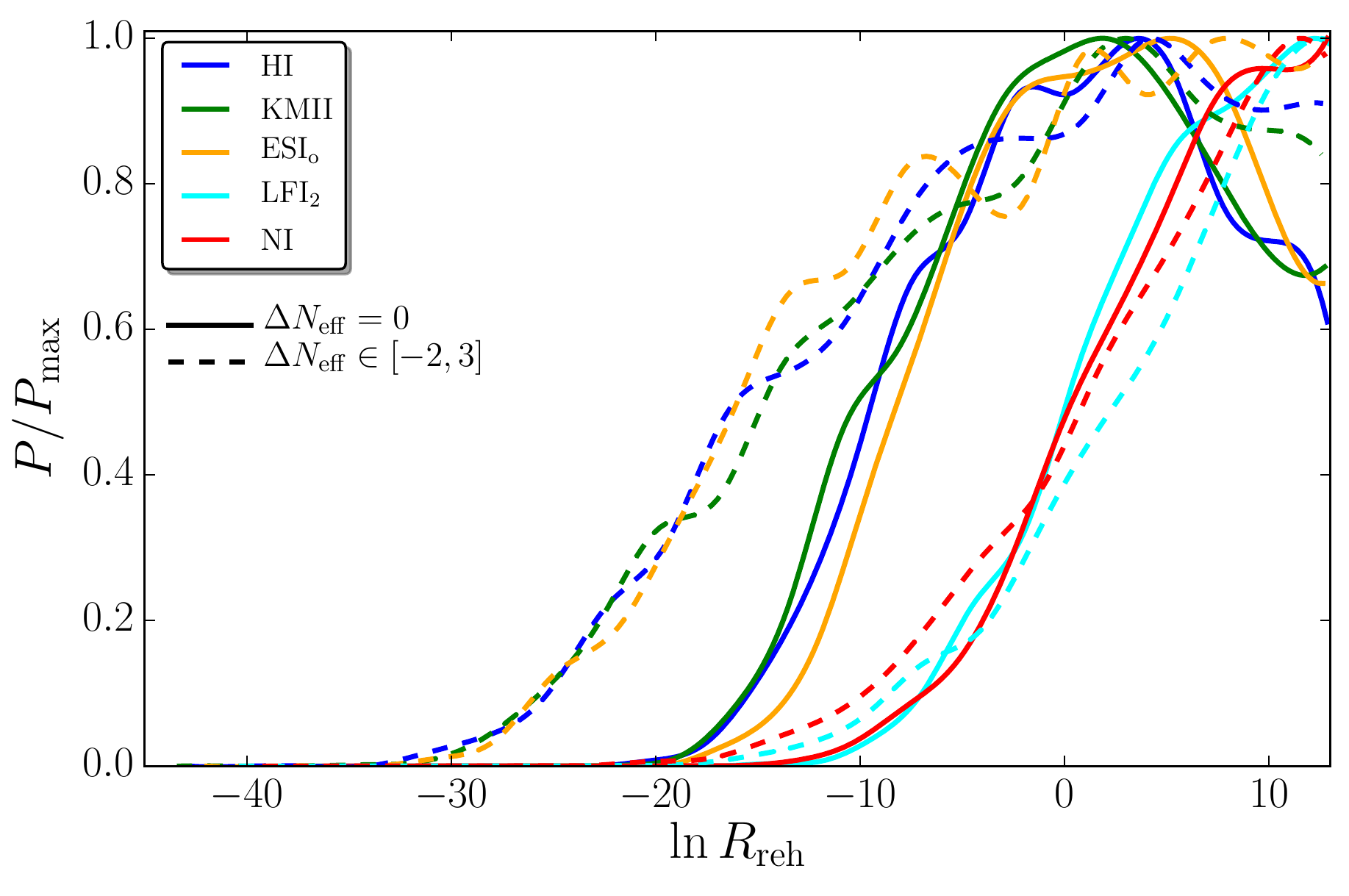}
\caption{Posterior distributions on the reheating parameter $R_\ureh$ for the single-field models considered in this paper, in the standard cosmological model (solid lines) and when $N_\ueff$ is allowed to vary (dashed lines). The analysis is performed with the Planck 2015 lowP + TT + lensing data. Power-law inflation is not displayed since this potential is conformally invariant. This means that its predictions do not depend on $R_\ureh$ and flat posterior distributions would be obtained in both cases.}
\label{fig:Rreh}
\end{center}
\end{figure}
As mentioned in \Sec{sec:intro}, the amount of expansion realised during reheating determines the amount of expansion realised between the Hubble exit time of the scales probed in the CMB and the end of inflation, hence the location of the observational window along the inflationary potential. More precisely, the number of \efolds elapsed between Hubble exit time of the pivot scale $k_{{}_\mathrm{P}}$ and the end of inflation is given by~\cite{Martin:2006rs, Martin:2010kz, Easther:2011yq} 
\bea
\Delta N_* = \ln R_\ureh + \frac{1}{2}\ln\left(\frac{\rho_*}{3\rho_\uend}\right)-\ln\left(\frac{k_{{}_\mathrm{P}}/a_\mathrm{today}}{\tilde{\rho}_{\gamma,\mathrm{today}}^\frac{1}{4}}\right)\, ,
\eea
where $\rho_*$ is the energy density at Hubble exit time of $k_{{}_\mathrm{P}}$, $\rho_\uend$ is the energy density at the end of inflation, $a_\mathrm{today}$ is the present value of the scale factor, $\tilde{\rho}_{\gamma,\mathrm{today}}$ is the energy density of radiation today rescaled by the number of relativistic degrees of freedom, and $R_\ureh$ is the reheating parameter, defined as
\bea
\ln R_\ureh & = 
\displaystyle
\frac{1-3\bar{w}_\ureh}{12\left(1+\bar{w}_\ureh\right)}\ln\left(\frac{\rho_\ureh}{\rho_\uend}\right)+\frac{1}{4}\ln\left(\frac{\rho_\uend}{\Mp^4}\right)\, .
\eea
In this expression, $\rho_\ureh$ is the energy density at the end of reheating, \ie at the onset of the radiation dominated epoch, and $\bar{w}_\ureh\equiv \int w(N)\dd N/N_\ureh$ is the averaged equation of state parameter during reheating. 

For a given inflationary potential, cosmological measurements therefore allow us to constrain the combination of $\rho_\ureh$ and $\bar{w}_\ureh$ appearing in the reheating parameter $R_\ureh$. In \Fig{fig:Rreh}, the posterior distribution on $R_\ureh$ is given for the single-field models considered in this work. The case of power-law inflation is not displayed since this potential is conformally invariant and its predictions do not depend on $\Delta N_*$, and by extension $R_\ureh$. The reheating parameter is thus left unconstrained in this model. For $\mathrm{LFI}_2$ and $\mathrm{NI}$, the constraints on $R_\ureh$ do not vary much when $N_\ueff$ is allowed to vary. For plateau potentials however ($\mathrm{HI}$, $\mathrm{KMII}$ and $\mathrm{ESI}_\mathrm{o}$), slightly lower values of $R_\ureh$ are allowed. This corresponds to lower values of $\nS$ at low $r$ which are more easily accommodated in the dark radiation extension according to \Fig{fig:nsr}. But beyond this difference, we find that, as for the Bayesian evidence of the models themselves in \Sec{sec:modelSelection}, constraints on the reheating expansion history are rather robust under the introduction of extra relativistic species.
\subsection{Constraining the effective number of relativistic species}
\label{sec:Neff}
\begin{figure}[t]
\begin{center}
\includegraphics[width=0.7\textwidth]{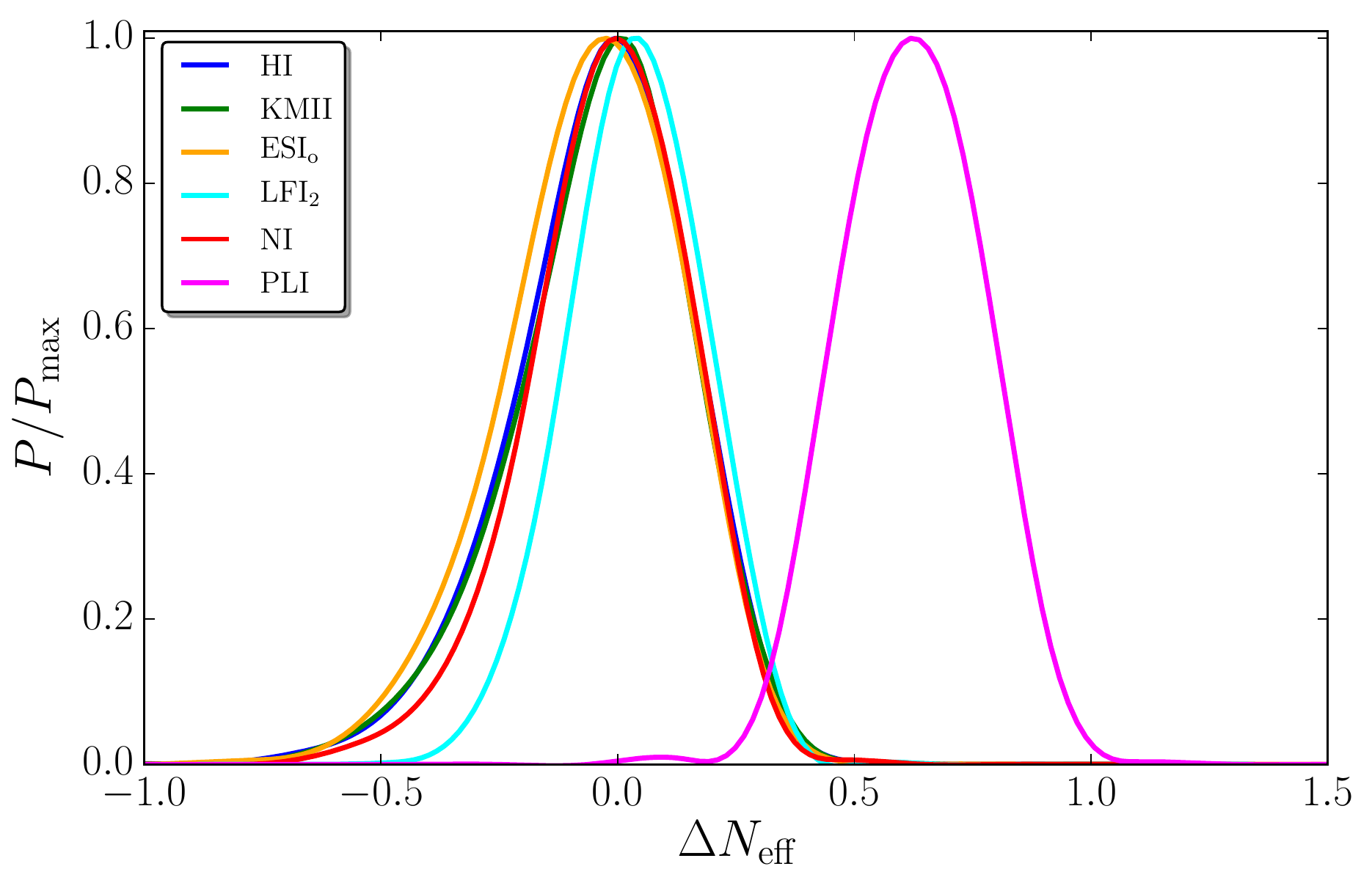}
\caption{Posterior distributions on the effective number of neutrino species $\Delta N_\ueff = N_\ueff-3.046$ for the single-field models discussed in this paper, from the Planck 2015 lowP + TT + lensing data.}
\label{fig:Neff}
\end{center}
\end{figure}
In \Sec{sec:modelSelection}, it was shown that even if the observational status of most inflationary models is unchanged when $N_\ueff$ is allowed to vary, disfavoured potentials yielding values of $\nS$ that are too large in the standard cosmological scenario, such as power-law inflation, can be brought back into the favoured zone once extra dark radiation components are included. This should lead to different predictions for $\Delta N_\ueff$.

In \Fig{fig:Neff} we show the posterior distributions of $\Delta N_\ueff$. The posteriors for all other potentials than $\mathrm{PLI}$ are almost the same and the mean value is the standard value $\Delta N_\ueff\simeq 0$. For $\mathrm{PLI}$ however, larger values of $N_\ueff$ are strongly preferred as expected, and we find 
\bea
\Delta N_\ueff^\mathrm{PLI} = 0.62^{+0.169}_{-0.168}
\eea 
where the bounds are $1\sigma$. This value would be consistent with a partly thermalised sterile neutrino or a Goldstone boson~\cite{Weinberg:2013kea}. 

\subsection{Including local measurements of the present expansion rate}
\label{sec:H_0}
\begin{figure}[t]
\begin{center}
\includegraphics[width=0.7\textwidth]{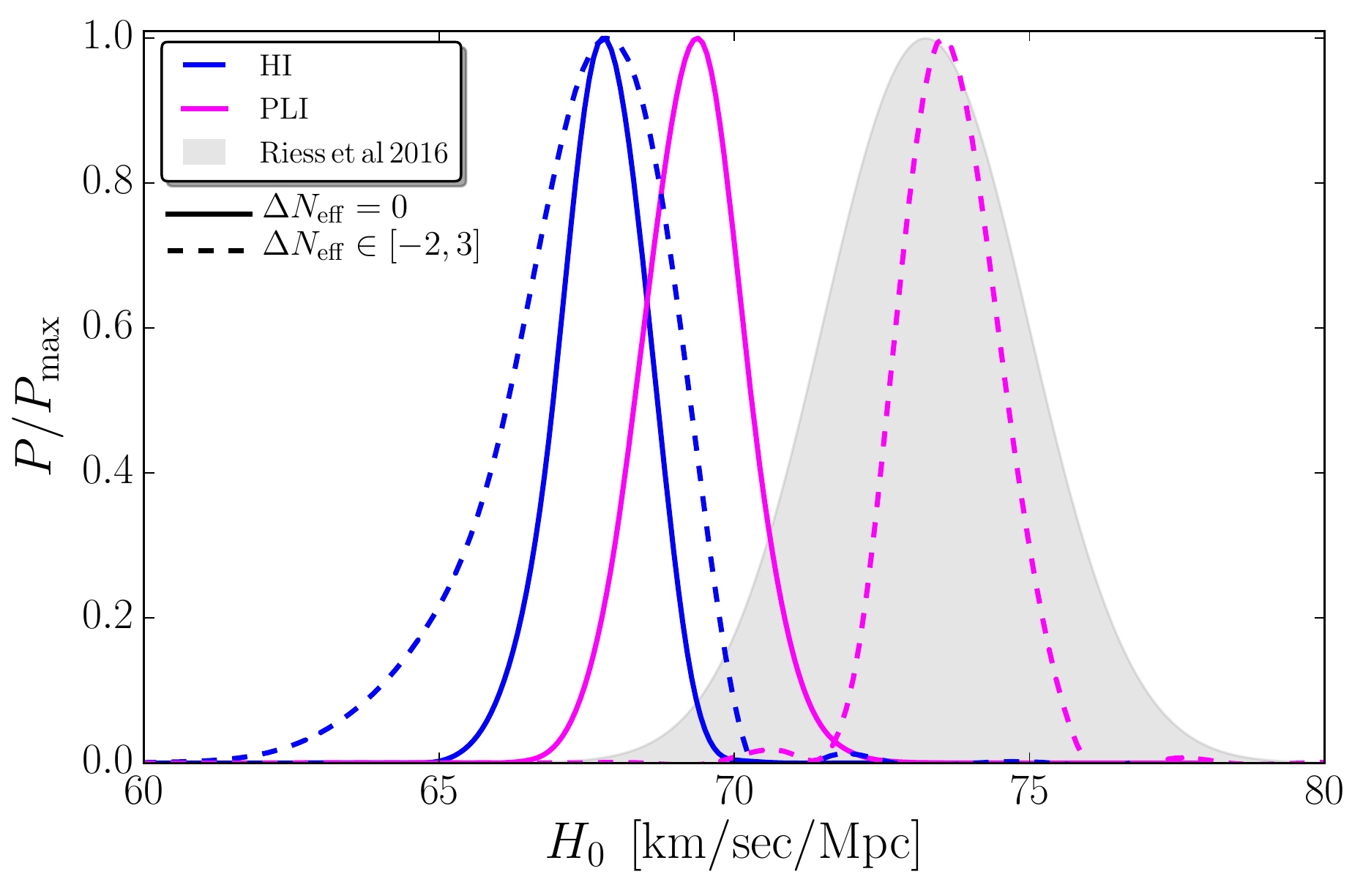}
\caption{Posterior distributions on the Hubble factor today $H_0$ for Higgs inflation (HI) as representative of plateau potentials and power-law inflation (PLI), with and without extra dark radiation. For PLI without dark radiation, the model is strongly disfavoured so the posterior distribution on $H_0$ is given for indicative purpose only. The grey area denotes the recent $H_0$ measurement by the Hubble Space Telescope presented in \Ref{Riess:2016jrr}. These posteriors are derived with the Planck 2015 lowP + TT + lensing data.}
\label{fig:H0}
\end{center}
\end{figure}

Dark radiation not only affects the growth of perturbations but also plays a role at the background level through its contribution to the expansion rate. This is why the present Hubble scale $H_0$ is relevant when discussing the degeneracy between $\nS$ and $N_\ueff$~\cite{2013PhRvD..87h3008H, 2016arXiv160300391G}. In \Fig{fig:H0}, the posterior distribution on $H_0$ has been displayed for $\mathrm{HI}$,  representative of plateau potentials, and $\mathrm{PLI}$, with and without dark radiation. The grey area represents the recent $H_0$ measurements~\cite{Riess:2016jrr} by the Hubble Space Telescope (HST). If inflation is realised with a plateau potential, there is some tension~\cite{Efstathiou:2013via} between CMB anisotropy data and the local HST measurement even when including dark radiation. However, power-law inflation yields a different value of $H_0=73.6 \pm 0.95\, \text{km}/\text{sec}/\text{Mpc}$ when including dark radiation due to the correlation between $\Delta N_\ueff$ and $H_0$.

\begin{figure}[t]
\begin{center}
\includegraphics[width=0.495\textwidth]{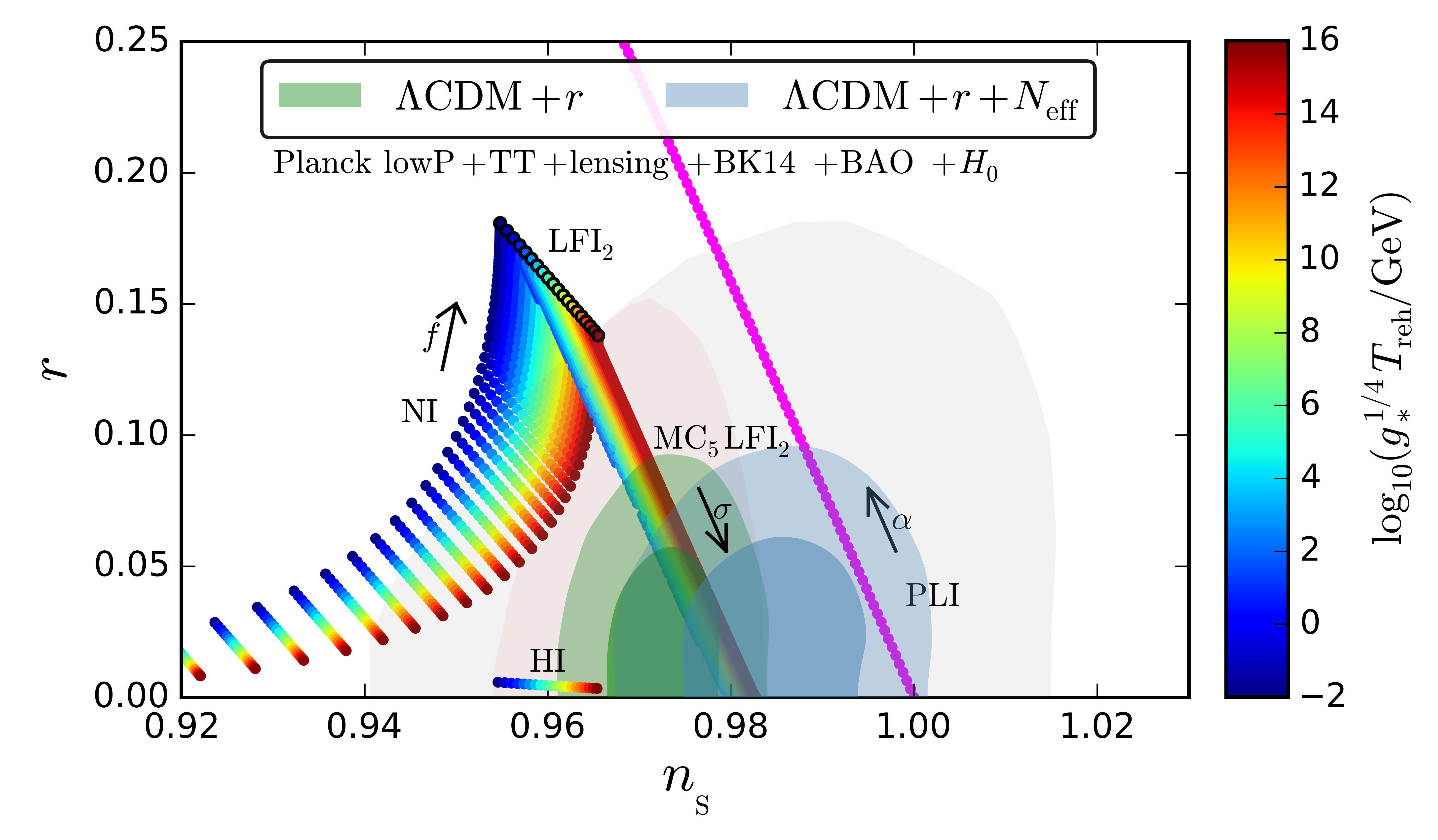}
\includegraphics[width=0.495\textwidth]{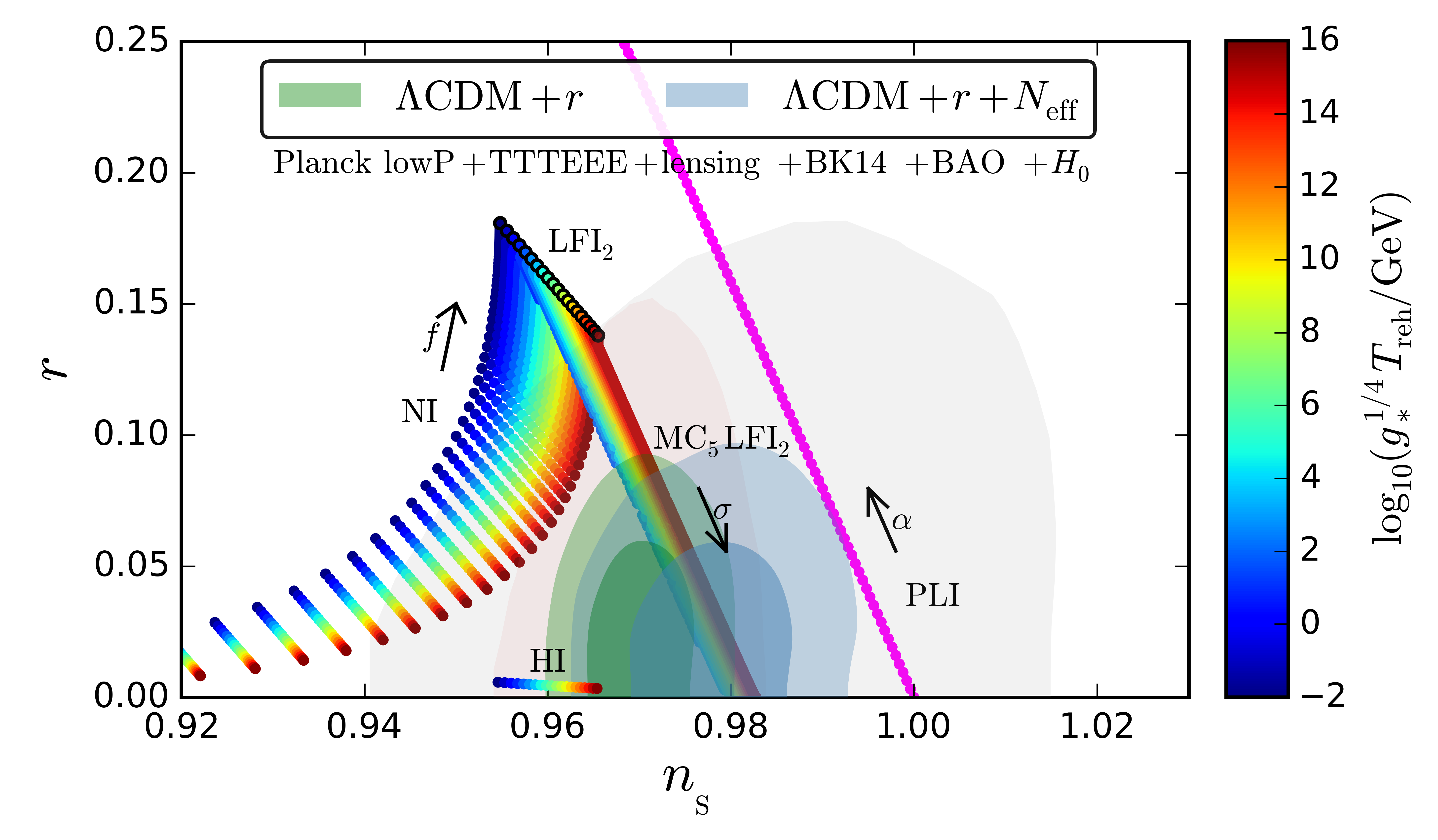}
\caption{Same as in \Fig{fig:nsr:datasets} with observational constraints derived when adding the local measurements on $H_0$. The green shaded surfaces are the one and two sigma contours when the standard $\Lambda\mathrm{CDM}$ cosmological model is assumed, while for the blue shaded surfaces, the effective number of relativistic degrees of freedom $N_\ueff$ is allowed to vary. In both panels, the constraints from the base Planck 2015 lowP+TT+lensing data displayed in \Fig{fig:nsr} have been recalled (pink and grey shaded areas).}
\label{fig:nsr:H0}
\end{center}
\end{figure}

The value obtained for $H_0$ in power-law inflation in a dark radiation cosmology is compatible with the recent Riess et. al. measurement~\cite{Riess:2016jrr} of $H_0=73.24\pm 1.74\, \text{km}/\text{sec}/\text{Mpc}$. This suggests that, by including local measurements of $H_0$ in the analysis, the status of power-law inflation compared to plateau potentials might change even when the more complete data sets discussed in \Sec{sec:datasets} are included. For this reason, in \Fig{fig:nsr:H0}, the posterior distributions on $\nS$ and $r$ are derived with the same data sets as in \Fig{fig:nsr:datasets} but further adding local measurements of $H_0$. In the case where  high-$\ell$ polarisation is not included (see \Sec{sec:datasets} for caveats on using this extended data set), one can check that the addition of local $H_0$ constraints allows larger values of $\nS$ to be favoured again when $N_\ueff$ is free to vary. This translates into improved Bayesian evidence for $\mathrm{PLI}$ since one finds $\ln\left(\mathcal{E}^\mathrm{PLI}/\mathcal{E}^\mathrm{HI}\right)=-14.50$ when $\Delta N_\ueff=0$ but $\ln\left(\mathcal{E}^\mathrm{PLI}/\mathcal{E}^\mathrm{HI}\right)=1.73$ when $N_\ueff$ is free to vary. In this case, $\mathrm{PLI}$ becomes one of the best single-field models of inflation.
\section{Conclusion}
\label{sec:Conclusions}

In this paper, we have investigated how much the standard ranking of inflationary models depends on the underlying cosmology. We have carried out a Bayesian model comparison for a few representative inflationary potentials, in the standard cosmology and when extra dark radiation is included. We have found that the observational status of most inflationary models is unchanged and that this ranking is fairly robust under the introduction of dark radiation. An exception is potentials such as power-law inflation that yield too large values of the scalar spectral index in the standard cosmological model. In the extended cosmology, power-law inflation belongs to the favoured group of single-field slow-roll models, but it implies that $\Delta N_\ueff=0.62^{+0.169}_{-0.168}$.

We also considered curvaton scenarios as examples of non single-field slow-roll models. We have shown that quadratic inflation with a curvaton, which is disfavoured in the standard cosmological model, becomes favoured if extra dark radiation is allowed. Finally, constraints on the reheating expansion history have been shown to be rather robust under the introduction of dark radiation components.

When the Bicep2/Keck + BAO data is included in the analysis, we have found that power-law inflation becomes strongly disfavoured again, but that the further addition of local measurements of $H_0$ makes it one of the best models of single-field slow-roll inflation. In this case, one would obtain a clear indication for the existence of an extra dark radiation component, with $\Delta N_\ueff\sim 0.62$. Therefore, even though the status of power-law inflation is still reliant on the way cosmological data sets are combined, this model seems to remain an interesting inflationary candidate that may shed new light on the physics of dark radiation.
It also illustrates why if the $H_0$ tension is to be resolved by dark radiation, it will have profound consequences for inflation model selection.

\acknowledgments
This work is supported by STFC grants ST/K00090X/1 and ST/N000668/1. It was initiated as a summer student placement project and RV thanks the Ogden Trust for support. Numerical computations were done on the Sciama High Performance Compute (HPC) cluster which is supported by  the ICG, SEPNet and the University of Portsmouth.

\subsubsection*{Note added}
After completion of this work, \Ref{DiValentino:2016ucb} investigated whether power-law inflation in the presence of dark radiation can also solve another tension in the standard cosmological model, namely the one on $\sigma_8$ between the Planck data and the weak lensing measurements. Their results seem to indicate that it is not the case.

\bibliographystyle{JHEP}
\bibliography{neutrinf}
\end{document}

%% file: newcommands.tex
\newcommand{\ie}{{i.e.~}}

\newcommand{\eg}{\textsl{e.g.~}}

\newcommand{\dd}{\mathrm{d}}
\newcommand{\ee}{e}

\newcommand{\sss}[1]{{\scriptscriptstyle{#1}}}

\newcommand{\uPl}{\mathrm{Pl}}

\newcommand{\uend}{\mathrm{end}}

\newcommand{\ueff}{\mathrm{eff}}

\newcommand{\ureh}{\mathrm{reh}}

\newcommand{\uS}{\mathrm{S}}

\newcommand{\usssS}{\sss{\uS}}

\newcommand{\usssPl}{\sss{\uPl}}

\newcommand{\nS}{n_\usssS}

\newcommand{\calP}{\mathcal{P}}

\newcommand{\GeV}{\mathrm{GeV}}

\newcommand{\Mpc}{\mathrm{Mpc}}

\newcommand{\Mp}{M_\usssPl}

\newcommand{\efolds}{$e$-folds~}

\newcommand{\beq}{\begin{equation}}
\newcommand{\eeq}{\end{equation}}
\newcommand{\bea}{\begin{eqnarray}}
\newcommand{\eea}{\end{eqnarray}}

\newlength{\wsingfig}
\newlength{\wdblefig}
\newlength{\wquadfig}
\newlength{\wtriplefig}
\newcommand{\Eq}[1]{equation.~(\ref{#1})}

\newcommand{\Fig}[1]{figure~{\ref{#1}}}

\newcommand{\Ref}[1]{ref.~{\cite{#1}}}

\newcommand{\Sec}[1]{section~\ref{#1}}
\newcommand{\Secs}[1]{sections~\ref{#1}}